# On the upper limit of percolation threshold in square lattice

Yu.P.Virchenko, Yu.A.Tolmacheva

The problem of percolation along sites of square lattice is studied. The number of contours being external boundaries for finite clusters has been estimated using geometric considerations. This estimation makes it possible to determine more precisely the upper limit for the percolation threshold.

Many problems of statistical theory of phase transitions are reduced to discrete problems of percolation theory. Although the formulation of mathematical problems of that theory seems to be rather simple and the physical essence of phenomena described by its models is clear, no substantial advance has been achieved to date in the problem of phase diagram calculations (in one-dimensional case, the percolation threshold). This situation is characteristic incidentally for the whole statistical theory of phase transitions. The solution of problem of percolation theory can be obviously described by construction of sequential approximations $Q^{(n)}(c)$, $n = 0, 1, 2...$ for such percolation probabilities $Q(c)$ that, starting already from the zero approximation, those have a non-trivial boundary $c^{(n)}$ for the concentration value region dividing the area where $Q^{(n)}(c) > 0$ and that where $Q^{(n)}(c) = 0$. The $c^{(n)}$ converges to the limiting surface (in the single parameter case, to the percolation threshold $c^*$) that is the true boundary of the percolation area in the phase diagram.

This work provides a simplest improvement of the known upper estimation for the percolation threshold $c^*$ in the leakage problem in a square lattice. This estimation can be considered as the zero approximation $c^{(0)}$ to $c^*$ when constructing the sequential approximations.

Let $\{\tilde{c}(x); x \in Z^2\}$ be a random Bernoulli field at a square lattice with the state probability in each site $x$,

$$P\{\tilde{c}(x) = 1\} = c, \quad P\{\tilde{c}(x) = 0\} = 1 - c.$$

The parameter $1 \geq c \geq 0$ is referred to as concentration.

Such a field can be described in terms of random sets $\{\tilde{M}; \tilde{M} \subset Z^2\}$, $\tilde{M} = \{x; \tilde{c}(x) = 1\}$ that will be called configurations. Each configuration, $\tilde{M}$ is characterized uniquely by a family of linked non-intersecting components $\tilde{W}_j$, $j = 1, 2, ...$ i.e., clusters wherein it is resolved, $\tilde{M} = \bigcup_j \tilde{W}_j$.

Let the probability $Q(x,c)$ of existence of an infinite cluster containing the site $x$, be of interest. If $Q(x,c) \neq 0$, one can state the existence of percolation at the field $\tilde{c}(x)$ from the site $x$. The calculation of the probability implies, in particular, the determination of so-called percolation threshold $c^*$ that is defined as

$$c^* \equiv \sup\{c : Q(x,c) \neq 0\}.$$

It is most likely impossible to calculate this characteristic for a square lattice. Therefore, only its approximate determination can be considered. To date, the percolation threshold $c^*$ has been proven to belong to the interval $[1/3, 6/7]$ [2]. In this work, the upper limit of the parameter $c^*$ will be estimated at a higher accuracy. Furthermore, the accuracy of the approximate $Q(x,c)$ probability calculation will be estimated.

Let the probability of percolation absence, $\overline{Q}(x,c) = 1 - Q(x,c)$ be considered. Let $A(x) = \{W : |W(x)| < \infty\}$ be the family of finite clusters $W(x)$ containing the site $x$. Let be introduced for any site $x \in Z^2$ as well as for any cluster the event $A(x,W)$ consisting in that the cluster containing the site $x$ coincides with $W$ within a random configuration $\tilde{M}$. Such an event has a certain probability equal to

$$\mathbf{P}\{A(x,W)\} = c^{|W|}(1-c)^{|\overline{W}|},$$

where the set $\overline{W}$ consists of lattice sites $y$ being adjacent to sites $x$ of the cluster $W$ but not belonging thereto in the sense of the connectedness relation $\varphi$ of the square lattice. Thus, $x$ and $y$ are connected together by one of relationships $y = x \pm e_1$, $y = x \pm e_2$, $e_1 = (0,1)$, $e_2 = (1,0)$. From here on, the symbol $|A|$ will denote the number of elements in the set $A$.

It is convenient for the further consideration to introduce the connectedness relation $\overline{\varphi}$ of the square lattice type with two diagonals. The sites will be referred to as $\overline{\varphi}$-connected ones if those are connected in the sense of $\varphi$ or if one of the following relationships is met:

$$y = x + e_1 \pm e_2, \quad y = x - e_1 \pm e_2.$$

A closed finite contour $\gamma$ connected in the sense of the $\overline{\varphi}$-connectedness and involving the site $x$ answers to each cluster of the $A(x)$ family. In this connection, the family $B(x)$ of all such contours $\gamma$ let be taken into consideration. For each contour $\gamma$ of $B(x)$, let the event $B(x,\gamma)$ be determined consisting in that the contour $\gamma$ is the outer boundary $\gamma(W)$ for a cluster $W$ of the $A(x)$ family. The set $\gamma(W)$ consists of sites $u$ belonging to the boundary of cluster $\tilde{W}$ and forming an infinite path $\alpha$ in the sense of $\varphi$-connectedness along the lattice sites, the $u$ being the sole site in $\alpha$ belonging to the conjuction $W \cup \tilde{W}$.

The event $B(x,\gamma)$ can be presented as countable conjuction

$$B(x,\gamma) = \bigcup_{W \in A(x): \gamma(W) = \gamma} A(x,W)$$

and thus has a certain probability $P(\gamma,x)$

$$P(\gamma,x) = \sum_{W \in A(x): \gamma(W) = \gamma} \mathbf{P}\{A(x,W)\} = \sum_{W \in A(x): \gamma(W) = \gamma} c^{|W|}(1-c)^{|\overline{W}|}. \quad (1)$$

It follows therefrom that

$$P(\gamma,x) = (1-c)^{|\gamma|} \sum_{W \in A(x): \gamma(W) = \gamma} c^{|W|}(1-c)^{|\overline{W}\setminus\gamma|}.$$

The family $A(x)$ is resolved into non-intersecting classes. The clusters $W \in A(x)$ having the same outer boundary belong to one of those classes. Then,

$$\bigcup_{W \in A(x)} \ldots = \bigcup_{\gamma \in B(x)} \bigcup_{W \in A(x): \gamma(W) = \gamma} \ldots . \quad (2)$$

From (1), (2), the decomposition follows

$$\overline{Q}(x,c) = \sum_{\gamma \in B(x)} P(\gamma,x).$$

Taking into account the latter equality ant the invariance $P(\gamma + a, x + a) = P(\gamma,x)$ with respect to shifts $a$ along the lattice, we obtain a formula serving as the base for the estimation to be sought

$$Q(c) = 1 - \sum_{\gamma \in B(0)} P(\gamma), \quad (3)$$

where $P(\gamma) = P(\gamma,0)$, $Q(c) = Q(0,c)$.

Now let the main result of the work be formulated.

Theorem. *The percolation threshold $c^*$ of a random Bernoulli field on a square lattice meets the inequality $1/3 < c^* \leq 4/5$ and at $c > 4/5$, the probability $Q(c)$ can be calculated at any guaranteed pre-specified accuracy.*

Proof. The inequality $1/3 < c^*$ has been stated before (see, e.g., [3]). To prove $c^* \leq 4/5$, let the following elementary estimation be used

$$P(\gamma) \leq (1-c)^{|\gamma|}.$$

Using this expression to establish the upper limit of the sum in (3)

$$\sum_{\gamma \in B(0)} P(\gamma) \leq \sum_{\gamma \in B(0)} (1-c)^{|\gamma|} = \sum_{k=4}^{\infty} (1-c)^k S_k, \quad (4)$$

where $S_k = |\{\gamma \in B(0): |\gamma| = k\}|$, $k \geq 4$ for all simple contours.

Let $x = 0$. Let us estimate the upper limit for $S_k$ that is independent of $x$. Let a path $\rho = (je_1, j = 0, 1, 2\ldots)$ be drawn from the site 0 into infinity. Then each contour $\gamma$ from $B(0)$ will intersect that path without fail at a certain site $z_\gamma$. Let the intersection site $\overline{z}_\gamma$ nearest to 0 one be selected. All contours of $B(0)$ are partitioned into non-intersecting classes such that all contours with the same site $\overline{z}_\gamma$ belong to same class. This partition into classes induces decomposition such that the set of contours of $B(0)$ becomes partitioned into corresponding classes of contours having length $k$. Let $l$ be the distance from 0 to the site $\overline{z}_\gamma$. Then, $l = 1, 2,\ldots, (k-1)$; the class consisting of contours intersecting $\rho$ at the distance $l$ let be denoted as $C_l$. In this case,

$$S_k = \sum_{l=1}^{k-1} |C_l|.$$

Let us fix the number $l$ and introduce the counter-clockwise orientation for each contour. Then, a site $x_1$ following $\bar{z}_\gamma$ in the contour $\gamma$ may be only one of the following ones:

$$\{\bar{z}_\gamma + e_1,\ \bar{z}_\gamma + e_1 + e_2,\ \bar{z}_\gamma + e_2,\ \bar{z}_\gamma - e_1 + e_2\}.$$

In this connection, contours belonging to one class $C_l$ are partitioned into non-intersecting classes $C_l^{(i)}$, $i = 1, 2, 3, 4$, depending on the selection of 1st site $x_1^{(l,i)}$ in the contour. Thus,

$$S_k = \sum_{l=1}^{k-1} \sum_{i=1}^{4} |C_l^{(i)}|. \tag{5}$$

It is seen from the Fig. (a, b) that at the each passing step of the contour consisting of the outer boundary sites, there are no more than 5 possible ways of its continuation. Therefore, the contour $\gamma$ of a length $k$ can be continued from the site $x_1^{(l,i)}$ in no more than $5^{k-2}$ ways, since after passing from the site $x_1^{(l,i)}$ and subsequent $k-2$ sites $x_2^{(l,i)},\ldots,x_{k-1}^{(l,i)}$, it is the $\bar{z}_\gamma$ site that shall be the last one therein.

Therefore, $|C_l^{(i)}| < 5^{k-2}$ and, proceeding from (5), we obtain

$$S_k \leq 4 \cdot 5^{(k-2)} \cdot (k-1). \tag{6}$$

Substituting (6) into (4) and making summation, we obtain

$$\sum_{\gamma \in B(0)} P(\gamma) < \infty, \quad c > 4/5.$$

This makes it possible to use the Borel-Kantelli lemma for the family of events $B(\gamma,0)$, $\gamma \in B(0)$. According to that lemma, the probability of an event consisting in the simultaneous realization of infinite set of events from said family is 0. Then, a certain maximum contour $\gamma \in B(0)$ exists at the probability 1. A site and infinite cluster containing that site can be found outside of that maximum contour. This results in that the probability $Q(c)$ is positive.

Finally, using (6), we find the upper estimate of the sum:

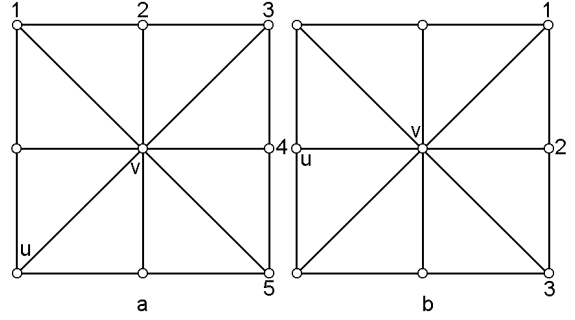

Fig. $u, v$, the sequential sites of the outer boundary $\gamma(W)$. Sites designated by numbers are possible continuations of the outer boundary contour.

$$\sum_{\gamma:\gamma\in B(0),|\gamma|\geq r} P(\gamma) \leq \sum_{\gamma:\gamma\in B(0),|\gamma|\geq r}(1-c)^{|\gamma|} = \sum_{k=r}^{\infty}(1-c)^k S_k =$$

$$= 4(1-c)^2 \sum_{k=r}^{\infty}(k-1)[5(1-c)]^{(k-2)} =$$

$$= 4(1-c)^2 \left(\frac{d}{d\zeta}\frac{\zeta^{r-1}}{1-\zeta}\right)_{\zeta=5(1-c)} < \infty.$$

The convergence is provided by the inequality $c > 4/5$. This estimate tends to zero at $r \to \infty$ and defines the accuracy of approximation for the percolation probability $Q(c)$ by the following polynomial in concentration, $c$:

$$Q(c) = 1 - \sum_{\gamma:|\gamma|<r,\ \gamma\in B(0)} P(\gamma).$$

To conclude, it is to note that the upper estimate of the percolation threshold $c^*$ obtained in this work is a rather rough one. It can be improved by excluding the self-intersections when estimating the number of contours, $S_k$. Nevertheless, principal importance represents the question whether such a procedure of sequential self-intersection exclusion would result in the accurate value of the threshold $c^*$.